\documentclass[twocolumn,showpacs,preprintnumbers,amsmath,amssymb,prl,aps]{revtex4-1}

\usepackage{graphicx}
\usepackage{dcolumn}
\usepackage{bm}

\begin{document}

\title{Fluorescence of laser created electron-hole plasma in graphene}

\author{Rainer J. St\"ohr$^{1,2}$, Roman Kolesov$^{1,2}$, Jens Pflaum$^3$, and J\"{o}rg Wrachtrup$^{1,2}$}
\affiliation{$^1$3. Physikalisches Institut, Universit\"at Stuttgart, 70550 Stuttgart, Germany\\$^2$Stuttgart Research Center of Photonic Engineering (SCoPE), 70569 Stuttgart, Germany\\$^3$Experimentelle Physik VI, Universit\"at W\"urzburg and ZAE Bayern, 97074 W\"urzburg, Germany}

\begin{abstract}
We present an experimental observation of non-linear up- and down-converted optical luminescence of graphene and thin graphite subject to picosecond infrared laser pulses. We show that the excitation yields to a high density electron-hole plasma in graphene. It is further shown that the excited charge carries can efficiently exchange energy due to scattering in momentum space. The recombination of the resulting non-equilibrium electron-hole pairs yields to the observed white light luminescence. Due to the scattering mechanism the power dependence of the luminescence is quadratic until it saturates for higher laser power. Studying the luminescence intensity as a function of layer thickness gives further insight into its nature and provides a new tool for substrate independent thickness determination of multilayer flakes.
\end{abstract}


\maketitle
\begin{text}
Due to its very unique properties graphene became a playground for studying fundamental aspects of relativistic charge carries confined in 2D \cite{Novoselov1, Novoselov2, Geim1}. This has led to insight into the electronic properties of graphene which is already far advanced \cite{Neto1}. However, the knowledge on its optical properties is still surprisingly narrow and mainly limited to reflection and absorption measurements \cite{Nair1, Stauber1, Ni1, Kumar1, Sun2}, transmission studies \cite{Dawlaty1, Sun1} or Raman spectroscopy \cite{Das1, Calizo1, Berciaud1, Ferrari2}. Here we report on a spectrally broad (several hundred nanometers) non-linear fluorescence of pristine graphene subject to picosecond laser irradiation. The qualitative model we provide is based on a high density electron-hole plasma where charge carriers can effectively gain or lose energy upon collision. The recombination of this plasma then leads to the fluoresence which is blue- and red-shifted with respect to the excitation wavelength. We further show that the blue-shifted part of the luminescence can be used for high-resolution, high-contrast imaging and thickness determination of single- and multilayer graphene flakes. We anticipate our work to be the starting point of a new field of graphene-related research combining the electronic properties of graphene with the fields of nano-photonics and relativistic plasma physics giving insight into dynamics of relativistic 2D electron-hole plasma.\\
Since its experimental realization \cite{Novoselov2} much attention has been devoted to graphene from both physics and device research communities. This has resulted in many intriguing experiments ranging for example from the anomalous quantum Hall effect \cite{Novoselov1, Novoselov3, Zhang1} to the Klein paradox \cite{Katsnelson1} and phase-coherent transport \cite{Miao1}. At the same time, optoelectronic applications like graphene based photodetectors are a newly emerging field \cite{Mueller1}. However, despite the fact that other low-dimensional carbon allotropes like carbon nanotubes \cite{Kim2, Liu1} or fullerenes \cite{Blau1, Yang1} show fascinating non-linear optical properties this area so far has remained unexplored in the case of graphene. While Raman studies or reflection measurements prove to be valuable for imaging and identifying appropriate flakes they do not reveal insight into the inherent luminescence properties of graphene. To our knowledge, so far fluorescence from individual graphene layers was only seen for example after oxygen treatment \cite{Gokus1} but not for pristine graphene. Here, we present intrinsic fluorescence from pure graphene. Surprisingly, fluorescence based imaging proved to be far superior to existing methods like Rayleigh imaging \cite{Casiraghi1}, reflection and contrast spectroscopy \cite{Ni1}, Raman spectroscopy \cite{Graf1}, or fluorescence quenching microscopy \cite{Kim1} due to the absence of fluorescence background from the substrate and impurities.\\
Graphene flakes were prepared by mechanical exfoliation of highly oriented pyrolytic graphite (HOPG) \cite{Novoselov2, Novoselov1} and subsequently transferred to suitable substrates. The two substrates mostly used in our studies were glass (150 $\mu$m thickness) and $\text{Si/Si}_3\text{N}_4$ (100 nm of nitride layer thickness). However, the same type of measurements showing similar results were performed with graphene on diamond, Si/SiO$_2$ (300 nm of dioxide layer thickness), and sapphire substrates. The output of a Kerr-lens mode-locked Ti:Sapphire laser was used for luminescence excitation after being passed through a single-mode photonic crystal fiber resulting in a total pulse length of 7 ps and a spectral width of 20 nm. The excitation wavelength could be tuned from 800 nm to 900 nm. The wavelength mostly used in this studies was 820 nm, however other wavelength showed similar results. Confocal measurements were carried out in a home-build confocal microscope. The parallel laser beam was focused onto the sample by using either a 1.3 NA oil immersion objective or a 0.85 NA air objective resulting in a diffraction limited spot. The sample was mounted on a 3D nano-positioning stage. The fluorescence light was separated from the laser light by a 50/50 beam splitter and send through a 100 $\mu$m pinhole for spatial filtering which allows only the in-focus portion of the light to be detected. Images were recorded by an avalanche photodiode (APD) with the light spectrally filtered by a 100 nm wide bandpass fiter centered around a wavelength of 700 nm. Spectra were acquired by a grating spectrometer equipped with a cooled CCD camera after blocking the excitation light by either a 780 nm shortpass or a 905 nm longpass filter.
\begin{figure}
\includegraphics[width=0.5\textwidth]{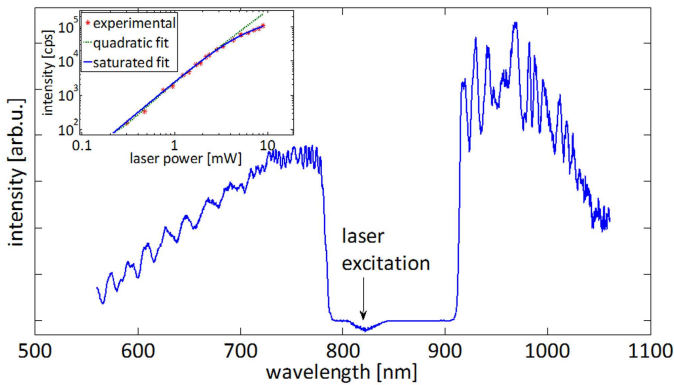}
\caption{\label{fig_1} Emission spectra of a single graphene layer under picosecond excitation of 820 nm wavelength. The periodic modulation on top of the spectrum is due to multiple reflections inside the optical elements of the setup and can therefore be considered an artifact. The gap from 800 nm to 900 nm is due to optical filters blocking the excitation light. The inset shows the dependence on laser power.}
\end{figure}
The spectrum of the observed graphene luminescence is shown in Figure \ref{fig_1} together with its dependence on incident laser power. In these measurements, a diffraction limited pulsed laser beam ($\lambda$ = 820 nm, $\tau$ = 7 ps) is optically exciting the graphene while the luminescence is detected by a grating spectrometer (For further details see experimental section). The graphene emission is a spectrally broad luminescence which peaks at the laser excitation wavelength and extends a few hundreds of nanometers in both the blue- and the red-shifted spectral region. 
To clarify the origin of this emission, we studied the influence of several key parameters. The intensity of the luminescence is dependent on the number of graphene layers and is roughly proportional to it for thin flakes. This suggests that no photochemical reactions due to surface states are involved as otherwise only the top layer would have contributed to the emission. This is also confirmed by the fact that the emission spectrum is independent of the substrate material used on either side of the graphene. The substrates of this study include diamond, sapphire, SiO$_2$, Si$_3$N$_4$ and glass examinated by oil and air objectives. Also, no change on the spectrum could be observed upon cooling to 4 K. The effects of rapid local heating of graphene by the laser pulse can also be disregarded since the black-body radiation would correspond to local temperatures on the order of a few thousand Kelvin and, consequently, burning graphene in contact with air. Furthermore, spectra were taken for different excitation pulse energies ranging from 2 pJ to 40 pJ, but no spectral shift of the luminescence was observed. At the same time, the absence of the defect induced D-line in the Raman spectrum after illumination indicates that no damage was inflicted onto the flake. The fact that the emission lifetime appears to be shorter than the time resolution of the setup (1 ns) supports the idea that emission is caused by ultra-fast electronic processes. Finally, we considered the influence of incident laser power density on the intensity of the emission. While for lower power densities $P$ the intensity depends quadratic on laser power (green line in Figure \ref{fig_1}) it saturates for higher power densities following a dependence like $I\propto\log(1+(P/I_0)^2)$ (blue line in Figure \ref{fig_1}).\\
It is known that optical absorption of graphene is $2.3\%$ in the wide range of frequencies from mid-IR to near-UV \cite{Nair1}. Each absorbed photon of frequency $\omega$ would produce an electron-hole pair with the energy being equally split between the electron and the hole. A 30 pJ laser pulse of 800 nm wavelength being focused into a nearly diffraction-limited spot of 500 nm in diameter would result in creation of $2.8\times 10^6$ relativistic mono-energetic electron-hole pairs corresponding to one pair per benzene ring. In such a high-density plasma electrons and holes would inevitably collide. The peculiarity of the situation is that electrons and holes interact via instantaneous Coulomb potential (probably, screened by plasma) even though they are relativistic, i.e. massless \cite{Neto1}. Neglecting screening effects, the classical Hamiltonian describing the interaction of two massless charges $q_{1,2}$ can be written as follows:
\begin{equation}
\label{eq:hamiltonian}
H=v_F\left(\left|\boldsymbol{p}_1\right|+\left|\boldsymbol{p}_2\right|\right)+\frac{q_1 q_2}{\left|\boldsymbol{x}_1-\boldsymbol{x}_2\right|},
\end{equation}
where $\boldsymbol{p}_{1,2}$ and $\boldsymbol{x}_{1,2}$ are the momenta and the positions of the corresponding particles and $v_F$ is the Fermi velocity. It is known from plasma physics, that energy exchange between scattering particles can be very efficient if they are having equal masses (electron-electron collisions), but very inefficient for particles of substantially different masses (electron-ion collisions). By analyzing the equations of motion deduced from Hamiltonian (\ref{eq:hamiltonian}), it can be shown that electrons and holes can efficiently gain or lose momentum and, consequently, energy even after a single collision (see Supplementary Information for more details). Thus, originally mono-energetic distribution of electrons and holes will diffuse in momentum and energy space as shown in Figure \ref{fig_2}. Finally, scattered electrons and holes can recombine giving rise to fluorescence shifted to the blue or to the red depending on the energy gain or loss. The described qualitative model explains all the above observations. The quadratic power dependence of the fluorescence follows from the fact that the number of scattering events is proportional to the number of scattering particles and the number of scatterers, i.e. to the square of the electron-hole plasma density. Saturation behavior at high laser power densities naturally follows from saturation of the excitation transition, when all the electrons of energy $-\hbar\omega/2$ are promoted from the valence band into the conduction band and the plasma density can no longer increase.
\begin{figure}
\includegraphics[width=0.5\textwidth]{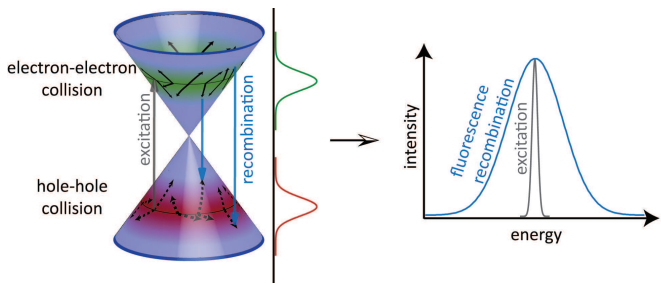}
\caption{\label{fig_2} Schematic representation of the 2D dispersion relation of graphene. The gray arrow shows the optical excitation of initially monoenergetic electron-hole pairs. Collisions lead to a broadening of the energy distribution as shown by the green and red curves on the right. Recombination of shifted electron hole pairs leads a broad fluorescence centered around the excitation energy.}
\end{figure}
Studying the dependence of this fluorescence on the number of layers yields further insight into its properties. In Figure \ref{fig_3} we compare the atomic force microscopy (AFM) image with the image taken by mapping of the blue-shifted part of the fluorescence of the same flake consisting of several different parts with thicknesses ranging from a single layer to approximately 20 layers. By this we can assign each layer a certain intensity of the upconverted fluorescence.
\begin{figure}
\includegraphics[width=0.5\textwidth]{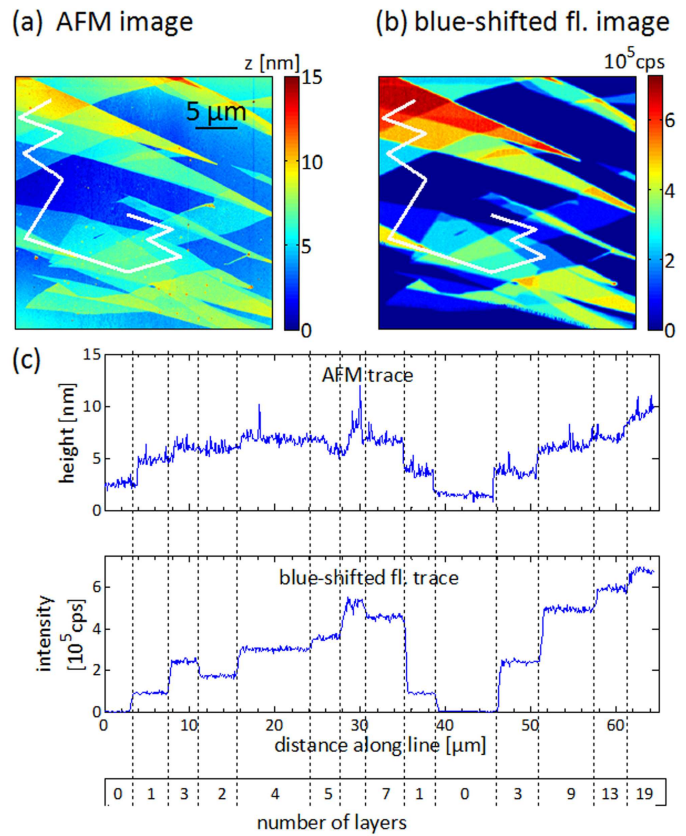}
\caption{\label{fig_3} Flake consisting of several parts of different thickness examined by AFM (a) and by imaging of the blue-shifted fluorescence signal (b). In (c) the profiles along the white lines are shown with the number of layers indicated at the bottom. The blue-shifted fluorescence was recorded in the spectral range from 650 nm to 750 nm. The substrate was Si/Si$_3$N$_4$.}
\end{figure}
To further understand this layer dependence we introduce a simple bulk model describing each component of the sample by its complex refractive index \cite{Roddaro1} (see inset of Figure \ref{fig_4} and Supplementary Information for details). Here, for simplicity we limit ourself to the case of graphene on glass substrate. However, we successfully modified the model to describe the thickness dependence of the luminescence for any dielectric substrate including Si$_3$N$_4$ or SiO$_2$ on silicon substrate. 
\begin{figure}
\includegraphics[width=0.5\textwidth]{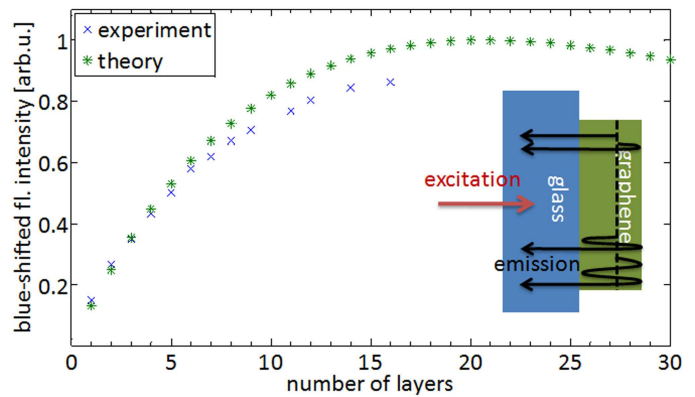}
\caption{\label{fig_4} Calculated dependence of the luminescence intensity on the number of graphene layers together with the experimental data. The inset shows a sketch of the underlying model. The red arrow symbolized the laser excitation while the black arrows show the luminescence contributions from a single graphene sheet taking into account multiple reflections at the different interfaces. We describe here the case of graphene on glass substrate however the model can be modified for any dielectric substrate.}
\end{figure}
The laser field distribution inside the flake is then being calculated by a transfer matrix formalism according to Fresnel law \cite{Roddaro1}. One thereby accounts for the 2.3\% absorption of each graphene layer \cite{Nair1} as well as for the interference of multiple reflections of the excitation light between the two interfaces. Both effects play a relevant role since the luminescence intensity is proportional to the fourth power of the electric laser field distribution. Subsequently, each graphene layer is treated as a individual source of luminescence and its contribution is calculated by summing up the Fabry-Perot type reflections from the different interfaces. Finally, the total luminescence intensity at the graphene-glass interface was calculated by integrating the intensity contributions from each individual layer. As can be seen from Figure \ref{fig_4}, the model reproduces the experimental dependence quite well, particularly for thin flakes. The deviation between the experimental and the theoretical values for thicker flakes is potentially due to saturation of the photodetector at higher count rates. It is important to note, that the model shown here is based on incoherent contributions from different layers. However, using a model with fixed phase relation between the emission and the excitation yields to a thickness evolution which is far from the experimental observations. This supports our interpretation of the fluorescence nature of the emission. Also, the fact that the model describes the experimental data well without taking into account interaction of plasma from different layers shows that the plasma is strongly confined within each layer.\\
The strong dependence of the luminescence on layer thickness together with its brightness make it a good candidate for imaging single and multilayer flakes. Particularly suitable for that purpose is the blue-shifted part of the luminescence since it combines the advantages of a second-order process with the absence of fluorescence background from the substrate or impurities. 
Figure \ref{fig_5} shows a flake imaged by the blue-shifted fluorescence and by standard optical wide-field microscopy for comparison. It already becomes obvious that particularly for very thin flakes like the single layer in the top right corner of Figure \ref{fig_5} the visibility is strongly improved in case of the blue-shifted luminescence. Also, the contrast between layers of different thickness is surprisingly high. Both of this aspects will now be discussed in more detail.
\begin{figure}[b]
\includegraphics[width=0.5\textwidth]{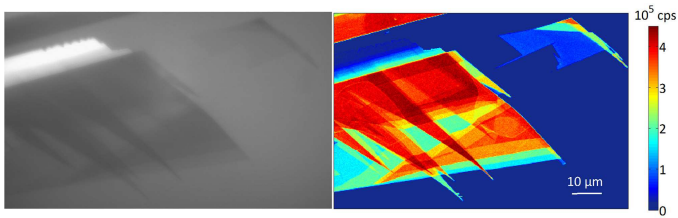}
\caption{\label{fig_5} Left: Optical wide-field microscopy image of a typical flake. Right: Imaging the same flake by mapping the fluorescence between 650 nm and 750 nm (excitation with 890 nm). The flake was prepared on a 150 $\mu$m thick glass substrate.}
\end{figure}
Based on the model presented above which contains no further fitting parameters it is possible to determine the numbers of layers for any arbitrary flake by comparing its intensity with the peak intensity for thicker layers. For example, it was found that the luminescence intensity from a single layer is 13\% of the peak intensity while a double layer emits 25\%. For comparison, the Raman 2D-line intensity of a single and a double layer is only 5\% and 6\% with noise of around 4\% and the ratio therefore significantly lower. This shows, that a major drawback of imaging graphene samples by observing the intensity of their Raman lines is the quite poor contrast between the bare surface and a single layer graphene due to rather low intensity of the Raman lines and strong background fluorescence \cite{Graf1}. In particular, imaging graphene by wide-field contrast and reflection microscopy is strongly limited since it also requires specific substrates \cite{Roddaro1}. However, in case of imaging graphene with the blue-shifted part of the fluorescence the situation is completely opposite: its occurrence is substrate independent, the signal is very strong and the background is very weak. This results in an image contrast defined by $C = (I_{graphene} - I_{substrate})/I_{substrate}$ of $C$ = 200 in case of the blue-shifted fluorescence while for Raman measurements at equal laser power and acquisition time the contrast is around 0.2. Finally, unlike AFM measurements, our non-invasive optical method does not cause any defects as confirmed by the absence of the Raman D line after imaging. \\
In conclusion, we experimentally demonstrated fluorescence of laser excited electron-hole plasma in graphene. Apart from studying interesting physics of high-density two-dimensional relativistic plasma, this fluorescence provides a very handy tool for visualizing graphene flakes and quantifying the number of layers in multilayered graphene. It adds another aspect to the arising graphene optical toolbox. As an outlook for future research related to this fluorescence, we would give a few examples, such that microscopy might enable one to follow the spatial dynamics of the laser induced plasma. Finally, graphene provides easy means to investigate the laser-induced plasma effects on its conductivity.\\

\begin{acknowledgments}
The authors gratefully thank Gopalakrishnan Balasubramanian, Petr Siyushev and Fedor Jelezko for helpful discussions and assistance.\\
Financial support was provided by the Landesstiftung BW ''Functional Nanostructures'' and ''Elitef\"orderung'' as well as the DFG via the research group FG730.
\end{acknowledgments}

\end{text}

\end{document}